# Multiple optical frequency-conversions in few-layer GaSe assisted by a photonic crystal cavity

*Liang Fang,¶ Qingchen Yuan,¶ Hanlin Fang, Xuetao Gan,\* Juntao Li, Tao Wang, Qinghua Zhao, Wanqi Jie, and Jianlin Zhao\**

L. Fang, Q. C. Yuan, Prof. X. T. Gan, Prof. J. L. Zhao
MOE Key Laboratory of Material Physics and Chemistry under Extraordinary Conditions, and Shaanxi Key Laboratory of Optical Information Technology, School of Science, Northwestern Polytechnical University, Xi'an 710072, China
E-mail: xuetaogan@nwpu.edu.cn; jlzhao@nwpu.edu.cn
Dr. H. L. Fang, Prof. J. T. Li
State Key Laboratory of Optoelectronic Materials Technologies, School of Physics, Sun Yat-sen University, Guangzhou 510275, China
Prof. T. Wang, Q. H. Zhao, Prof. W. Q. Jie
State Key Laboratory of Solidification Processing, Northwestern Polytechnical University, Xi'an 710072, China
¶ Contributed equally to this work.

While two-dimensional (2D) materials have intriguing second-order nonlinearities with ultrahigh coefficient and electrical tunability, their atomic layer thicknesses hinder explorations of other optical frequency-conversions (OFCs) than second harmonic generation (SHG) due to inefficient light-coupling and unachievable phase-matching. We report, by resonantly pumping a photonic crystal cavity integrated with a few-layer GaSe, it is possible to realize multiple second-order nonlinear processes in GaSe even with microwatts continuous wave pumps, including SHGs, sum-frequency generations (SFGs), cascaded SFGs and their induced third harmonic generations. These OFCs arise from the significant cavity-enhancements. The enhancement factor of a SHG process is estimated exceeding 1,300. The cascaded SFGs have comparably strong intensities with those of SHGs. To the best of our knowledge, this is the first observation of cascaded second-order nonlinear processes in 2D materials. The cavity-assisted OFCs could provide a view to study deterministic OFCs in 2D materials with low pump power and expand their optoelectronic applications to nonlinear regime.

## 1. Introduction

Optical frequency-conversions (OFCs) in bulky nonlinear materials have been recognized as a crucial route towards realizing novel desired coherent light sources, entangled photon-pairs, all-optical signal processing, etc.[1] To achieve a decent frequency-conversion efficiency, nonlinear materials with second-order nonlinear polarizability are employed prior over those with third-order nonlinearity.[2,3] Recently, two-dimensional (2D) materials with a thickness of few atomic layers have been revealed as a suitable candidate for OFCs, presenting a variety of advantages over bulk materials: (1) The low dimensionality of 2D materials gives rise to reduced dielectric screening of Coulomb interactions between charge carriers, allowing their strong light-matter interactions and exotic excitonic effects.[4,5] The second-order nonlinear susceptibility $\chi^{(2)}$ could be extraordinarily high. A value of $\chi^{(2)} \sim 1,000$ pm/V is estimated in a monolayer $WSe_2$, which is about three orders of magnitude higher than those in conventional bulk materials.[6,7] (2) Strong exciton charging effect in 2D materials allows for control over the oscillator strengths at the exciton and trion resonances, promising an electrical modulation of the exciton-enhanced SHG as well as the corresponding $\chi^{(2)}$ by one order of magnitude.[8] (3) $\chi^{(2)}$ is a third-rank tensor and therefore must vanish if the material has an inversion symmetry. When the layered materials are thinned down to 2D materials, the crystal structures are determined by the layer number strongly, leading to variations of the inversion symmetry. It is reported few-layer $MoS_2$ or BN with odd (even) numbers of layers present appreciable (negligible) second harmonic generations (SHGs), showing an obvious layer-dependent oscillation.[9,10] In addition, 2D materials' crystal structures could also be governed by their layer stacking orders or spiral angles,[11-14] which provides more degrees of freedom to engineer their second-order nonlinearity. (4) And besides, 2D materials' intriguing electrical properties and ability of chip-integration make them have benefits in constructing optoelectronic devices, such as high-performance photodetectors, modulators, and light-emitting diodes.[15-21] It would facilitate device applications of 2D materials' second-order nonlinearities.

While materials with $\chi^{(2)}$ could support a variety of OFCs, 2D materials' prominent second-order nonlinearities have only been illustrated by the realizations of SHGs. There are challenges to achieve other OFCs, such as sum-frequency generation (SFG), difference-frequency generation (DFG), and spontaneous down-conversion (SPDC). First, 2D materials are so thin that a normal incident light could not couple with them efficiently. Hence, in the reported SHGs of 2D materials, a pump of pulsed laser was essential to provide an extremely high peak power. This requirement makes it difficult to realize 2D materials' OFCs participated by two pump radiations, such as SFG and DFG. In those processes, the pulses of the two pump lasers should be coincident in their interactions with 2D materials to provide intense peak powers simultaneously.[22] Hence, a complicated experiment system to realize the synchronization of the two pulsed lasers is required. Second, in bulk materials, to increase efficiency of a certain OFC in the three-wave mixing, phase-matchings via angled input or periodic poling are utilized.[1-3] Unfortunately, 2D materials only have few atomic layers, it is impossible to design phase-matchings to strengthen specific OFC process over SHGs.

Here, we report, by integrating a few-layer GaSe onto a photonic crystal (PC) cavity, it is possible to realize multiple OFC processes. PC cavities fabricated on a semiconductor slab could provide resonant modes with high quality ($Q$) factors (>$10^3$) and cubic-wavelength mode volumes ($V_{mode}$), which therefore promise a high density of electrical field proportional to $Q/V_{mode}$. In our previous work, assisted by a PC cavity's resonant mode, high-efficiency SHGs in a few-layer GaSe were accomplished with few microwatts continuous wave (CW) pumps.[23] This exemption from the requirement of pulsed laser could facilitate simple implementations of 2D materials' OFCs by involving two CW pumps. In addition, these CW-pumped OFCs arise from the enhancements of specific cavity modes. It could be employed to determine the frequency-conversion channel and efficiency from the multiple possible three-wave mixing processes.

GaSe is a layered metal monochalcogenide crystal with two planes of Ga atoms sandwiched by two planes of Se atoms in each primitive layer. Unlike the well-studied layered transition metal dichalcogenides, the thinning of GaSe from bulk to few-layers will drive a direct-to-indirect bandgap transition, and the bandgap could be tuned reliably by controlling the layer numbers and doping levels in GaSe's van der Waals heterostructures.[24, 25] As a well-known nonlinear crystal over the spectral range from visible to terahertz, GaSe's monolayer has a second-order nonlinear coefficient about one order of magnitude larger than that of monolayer $MoS_2$.[14] If its layer stacking order has a $\varepsilon$-GaSe classification, which is employed in our work, the second-order nonlinearity has no layer dependence due to the absence of inversion center for arbitrary layer thicknesses. [26, 27] More importantly, it is possible to epitaxially grow large-area few-layer GaSe on semiconductor substrates, such as GaN and GaAs, [28, 29] which are normally used to fabricate high quality PC cavities. Hence, the layered GaSe is chosen as the 2D nonlinear material to integrate with the PC cavity in our experiment. We would note there are many other 2D materials that have potentials to integrate with PC cavities for low-power OFC operations due to their non-centrosymmetry and large-area growth on bulk semiconductor substrates, such as InSe, WSe$_2$, and their heterostructures. [30, 31]

From a PC cavity integrated with a five-layer GaSe, we observe clear multiple OFC processes by resonantly pumping cavity's two resonant modes, including SHGs, SFGs, cascaded SFGs, and third harmonic generations (THGs) induced by cascaded SFGs. With the cavity-enhancement factors around 1,000, the CW pump powers are greatly reduced to few microwatts. To the best of our knowledge, this is the first observation of cascaded second-order nonlinear processes in 2D materials. The obtained cascaded SFGs have comparable intensities with those of SHGs. The significantly cavity-enhanced OFCs in 2D materials, especially the possibility of utilizing CW lasers, promise the future studies of other three-wave mixings in 2D materials, such as DFG, SPDC, as well as the optical parametric oscillation and amplification.

## 2. Device Fabrication & Measurement Arrangement

The employed PC cavities are fabricated on a silicon-on-insulator wafer with a top 220 nm thick silicon slab (see Supporting Information). The cavity defect is formed by outward shifting the central four air-holes to endow an ultrasmall $V_{mode}$.[32] To facilitate its mode-coupling with the employed vertical microscope, part of the air-holes around the central defect are shrunk to shift the resonant mode into an asymmetric distribution.[33] The PC cavities are designed to have resonant modes at the telecom-band. Few-layer GaSe flakes are mechanically exfoliated from a bulk material on polydimethylsiloxane stamps,[34] which are then dry-transferred onto the PC cavity (see Supporting Information).

Figure 1(a) displays an optical microscope image of the integrated device. The cavity defect formed by shifted air-holes and periodically shrunk air-holes could be recognized around the PC center. The integration of few-layer GaSe modifies the color of the top silicon layer into light green. Atomic force microscopy (AFM) is used to distinguish GaSe's location on the PC cavity as well as its thickness, as shown in Fig. 1(b). The GaSe flake covers the cavity's defect precisely, promising effective interactions between the GaSe layer and cavity's modes. The measured thickness of 4 nm corresponds to a layer number of five by considering the monolayer thickness of 0.85 nm.[26] The observable PC air-holes under the GaSe flake indicate the conformal contact between GaSe and PC. In addition, there are some folded GaSe fringes caused by the imperfect transfer. Since these folded fringes locate outside the cavity defect region, they would not couple with the cavity mode effectively to generate different OFC responses, as illustrated in the spatial mappings of OFCs in the Supporting Information.

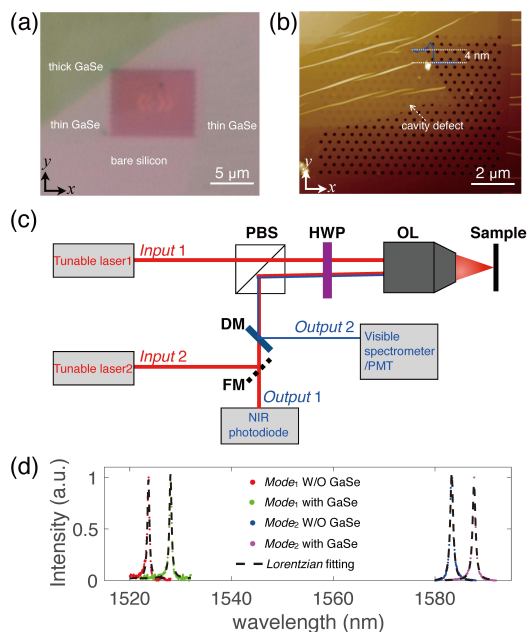

**Figure 1.** (a) Optical microscope image of the fabricated device, where the dark and light green regions indicate the thick and thin GaSe flakes, respectively. (b) AFM image of the fabricated device, showing a 4 nm thick GaSe flake over the cavity defect precisely. (c)

Schematic experiment setup for the cross-polarized reflection measurements and frequency-upconversion measurements; PBS: polarized beam splitter, HWP: half wave plate, OL: objective lens, DM: dichroic mirror, FM: flip mirror. (d) Reflection spectra of the PC cavity before and after the integration of GaSe, showing two red-shifted resonant modes.

Resonant modes of the integrated GaSe-PC cavity are characterized using a vertical cross-polarization microscope,[35] as schematically shown in Fig. 1(c). A telecom-band tunable laser as the excitation source incident from the *Input*1 port is focused on the PC cavity by a near-infrared (NIR) objective lens (OL). The resonant scattering of the PC cavity is collected by the same OL, and the optical power is monitored at the *Output*1 port by a sensitive NIR photodiode. The light paths of excitation laser and collection signal are separated by a polarized beam splitter (PBS) to ensure their orthogonal polarizations, which improves the signal to noise ratio of the resonant mode's reflection. A half-wave plate (HWP) is inserted between the PBS and OL. By rotating it, the polarization of the excitation laser could be controlled with respect to the cavity's orientation to guarantee an efficient mode-coupling. This cross-polarization microscope is also adopted to measure the OFCs from the GaSe-PC cavity. To do that, a longpass (cut-off wavelength of 1,000 nm) dichroic mirror (DM) is placed in the signal collection path to reflect the frequency-upconversion signal into a visible spectrometer or a photomultiplier tube at the *Output*2 port. To mix two pump lasers onto the GaSe-PC cavity, another tunable laser is incident from the *Input*2 port, which is controlled by a flip mirror (FM) in the signal collection path to replace the *Output*1 port.

## 3. Results & Discussions

Before and after the integration of the GaSe flake, we measure reflection spectra of the PC cavity by scanning wavelength of the tunable laser, as displayed in Fig. 1(d).

From the bare PC cavity, two resonant modes are obtained at the wavelengths of 1523.7 nm (*Mode*$_1$) and 1583.3 nm (*Mode*$_2$), and their *Q* factors are estimated as 3,100 and 2,200 by Lorentzian fittings, respectively. GaSe has a refractive index of ~2.8 and negligible absorption in the telecom-band spectral range. Its integration red-shifts the wavelengths of *Mode*$_1$ and *Mode*$_2$ to 1528.0 nm and 1587.6 nm, respectively, since the positive dielectric perturbation of the electromagnetic field. This result illustrates effective interactions between GaSe and resonant modes' evanescent field, which would facilitate cavity-enhanced OFCs. In addition, the ultrathin GaSe flake would not generate extra scattering loss of the resonant modes, as confirmed from the three-dimensional finite element simulations. Thence, there is no observable modification of the *Q* factors.

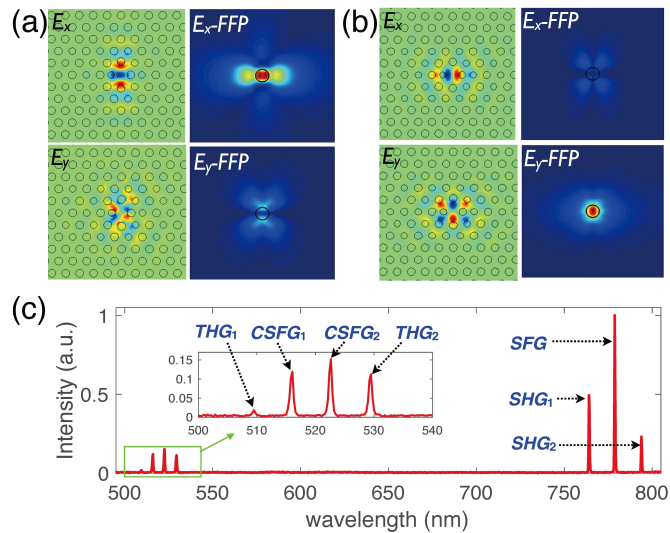

**Figure 2**. (a, b) Calculated near-field distributions and FFPs of the resonant *Mode*$_1$ (a) and *Mode*$_2$ (b). Black solid circles are included into the FFPs to show the collection region of the employed OL. (c) Frequency-upconversion spectrum obtained from the GaSe-PC cavity when pumped by two on-resonance lasers.

To realize cavity-enhanced OFCs in few-layer GaSe, it is essential to couple pump lasers into the cavity modes with high efficiencies. Considering the far-field coupling configuration of the employed microscope, it is worth calculating far-field patterns (FFPs) of the cavity's resonant modes. Figures 2(a) and (b) display the simulated near-field distributions ($E_x$ and $E_y$) and the corresponding FFPs of $Mode_1$ and $Mode_2$, respectively. FFPs are calculated from their near-fields using the Rayleigh-Sommerfeld diffraction integral theory.[36] In the displayed FFPs, black solid circles are added to indicate the amount of electrical fields that could be collected by the OL, which are defined by its numerical aperture. Because of the asymmetric near-field distribution, the FFP of $Mode_1$'s $E_x$ radiates vertically and concentrates mostly into a convergent region inside the OL's collection angle. For the $E_y$ component, the FFP orients to oblique directions and presents a split distribution. Most of the optical field distributes outside the black solid circle. Hence, to efficiently couple an incident laser into $Mode_1$ by the microscope, its polarization should be controlled along the *x*-direction. The calculated results shown in Fig. 2(b) illustrate $Mode_2$'s $E_y$ has a strongly localized FFP inside the collection angle, suggesting a high coupling efficiency for the *y*-polarized excitation. These polarization-dependent far-field couplings are verified experimentally by measuring the reflections of excitation lasers on-resonance with $Mode_1$ and $Mode_2$. Combining with the collection efficiency of 22% (39%) for $Mode_1$'s *x*-component ($Mode_1$'s *y*-component) calculated from Fig. 2(a) (Fig. 2(b)), the incident laser could couple into $Mode_1$ ($Mode_2$) with an efficiency estimated as 4.5% (6%).

In the OFC experiments, we first examine cavity-enhanced SHGs by individually coupling a CW laser on-resonance with $Mode_1$ (1528.0 nm) or $Mode_2$ (1587.6 nm) into the PC cavity. Thanks to the resonantly localized evanescent field of the PC cavity, strong SHGs are observed for both CW on-resonance pumps, even with powers of few microwatts. When the SHG conversion efficiencies are maximized, the incident polarizations of the two on-resonance lasers have a 90° difference, which is consistent with the above mode-coupling analysis. The factors of these cavity-enhanced SHGs could be estimated in the way demonstrated in our previous work.[23] Because of various optical losses in the microscope setup, it is complicated to measure the absolute SHG power. The on-resonance CW laser is switched into a pulsed laser, whose wavelength is off-resonant from the cavity modes. In this regime, both the on-resonance CW pump and the off-resonance pulsed pump share the same excitation and collection optical paths, as well as the same dielectric environment of the SHG emission. By comparing SHGs pumped by the on-resonance CW laser and an off-resonance pulsed laser, the enhancement factors by $Mode_1$ and $Mode_2$ are estimated as 1,320 and 714 (see Supporting Information), respectively.[23] These values are larger than the result reported in Ref. [23] due to the larger $Q$ factors of the employed resonant modes.

The two CW lasers on-resonance with $Mode_1$ and $Mode_2$ are then coupled into the PC cavity together. Considering the polarization-dependences of modes' far-field couplings, we input lasers at 1528.0 nm and 1587.6 nm from the $Input$1 and $Input$2 ports to match their orthogonal polarizations. Figure 2(c) plots the measured spectrum of the frequency-upconversion signals with the two pump powers measured around 100 μW after the OL. Considering the far-field coupling efficiencies of $Mode_1$ and $Mode_2$, the pump powers coupled into the resonant modes are less than 10 μW. Multiple OFC peaks are obtained in the measured spectrum. Around the wavelength of 780 nm, SHG peaks of the two pump lasers are observed at 764 nm ($SHG_1$) and 793.8 nm ($SHG_2$). Between them, there is another peak at the wavelength of 778.9 nm, corresponding to the SFG signal of the two pump lasers according to the wavelength calculations.

In the shorter wavelength range of the OFC spectrum, there are other four weak peaks at 509.3 nm, 515.8 nm, 522.4 nm, and 529.2 nm, as shown in the zoomed spectrum of Fig. 2(c). Except the peak at 509.3 nm, other three peaks have comparably high intensities with that of $SHG_2$. The wavelengths of 509.3 nm and 529.2 nm are coincident with THG wavelengths of the two pump lasers. Before GaSe's integration, the same OFC experiments are carried out on the bare silicon PC cavity with the pumps of two on-resonance lasers. THGs of the two pump lasers could also be obtained due to silicon's third-order nonlinearity (see Supporting Information).[37] However, their strengths are almost two orders of magnitude weaker than those achieved in the GaSe-PC cavity. In the 4 nm GaSe flake, the optical power is less than 1% of that in the 220 nm silicon slab.[38] Even if GaSe has a comparable third-order nonlinear susceptibility with that of silicon, [2,3] the THGs generated by the GaSe flake should be much weaker than those generated by the silicon slab. Hence, the strong THG signals obtained from the GaSe-PC cavity cannot be resulted from the third-order nonlinearities in either silicon slab or GaSe flake.

Considering GaSe's high $\chi^{(2)}$ as well as strongly localized resonant modes in the PC cavity, we attribute the obtained strong THGs to the cavity-enhanced cascaded second-order nonlinear processes in GaSe[39], i.e., the cavity-enhanced $SHG_1$ ($SHG_2$) mixes with the fundamental wave of $Mode_1$ ($Mode_2$) via a cascaded SFG process and generates a signal at the wavelength of $THG_1$ ($THG_2$). The ability of cascaded SFG in GaSe-PC cavity could also reveal the origin of the peak at 515.8 nm (522.4 nm), which is the cascaded mixing of $SHG_1$ ($SHG_2$) and fundamental wave of $Mode_2$ ($Mode_1$). To help discussions, we label the peaks at 515.8 nm and 522.4 nm as $CSFG_1$ and $CSFG_2$, respectively. In the bare silicon PC cavity, these two peaks could not be observed even when the pump powers are increased by two orders of magnitude (see

Supporting Information), suggesting the necessities of GaSe's second-order nonlinearity and cascaded processes.

These achievements of OFCs rely on GaSe's second-order nonlinearity and its coupling with the evanescent field of the PC cavity mode. The employed $\varepsilon$-GaSe has second-order nonlinearity for arbitrary thicknesses due to the noncentrosymmetry.[26, 27] Hence, for GaSe flakes with varied thicknesses, the cavity-enhanced OFC should still be valid.[23] Note that, because the intensities of second-order nonlinear processes have quadratic or cubic dependences on GaSe's layer numbers,[27] the cavity-enhanced OFCs will have reduced intensities if the GaSe was thinned down to few-layer or monolayer. As demonstrated in our previous work,[23] the cavity-enhanced SHG of a monolayer GaSe was greatly reduced by a ratio of 75 compared to that of a nine-layer GaSe.

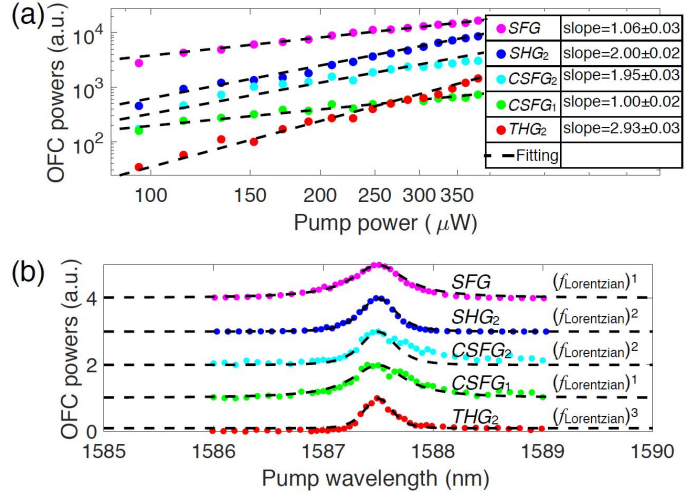

**Figure 3**. (a) Log-log plots of OFCs' power-dependences on the pump power of laser at 1587.6 nm, while the power of laser at 1528.0 nm is maintained. (b) Spectra of OFCs when the pump wavelength of laser at 1587.6 nm is scanned across $Mode_2$, where $f_{Lorentzian}$ is the Lorentzian function obtained from the fitting of $Mode_2$ in Fig. 1(d).

We recognize the above OFC processes by calculating their wavelength-conversions. To further verify them, we examine their pump power-dependences by varying the power of the laser at 1587.6 nm ($Mode_2$). The power of another pump at 1528.0 nm is maintained. The measured results are displayed in the log-log plots of Fig. 3(a). In the process of $SHG_2$ ($THG_2$), only the pump laser at 1587.6 nm participates, and two (three) photons of this fundamental wave convert into one photon of the frequency-upconversion wave. Hence, its signal power is a quadratic (cubic) function of the pump power, which agrees well with the polynomial fitting slope of 2±0.02 (2.93±0.03) from the experimental results. In other processes of $SFG$, $CSFG_1$ and $CSFG_2$, which involve both pump lasers, to generate one frequency-upconverted photon, the laser at 1587.6 nm contributes one photon, one photon, and two photons, respectively. Hence, the log-log plots of power-dependences for these three processes have slopes close to 1, 1, and 2, as indicated in Fig. 3(a).

The possibility of realizing microwatts-level CW pumped OFCs in the few-layer GaSe, specially the cascaded SFGs, arises from the cavity-enhanced light-matter interactions under the high density of electrical fields.[38,40] To prove that, we fix one of the pump lasers at the resonant wavelength of $Mode_1$ (1528.0 nm) and scan the other pump wavelength across $Mode_2$ (1587.6 nm). Figure 3(b) plots the obtained power variations of the five OFC processes participated by $Mode_2$. As the laser wavelength is detuned away from $Mode_2$, these OFC signals decrease to undetectable levels due to the weak light-GaSe coupling in the single-passed normal radiation. When the laser coupled into $Mode_2$ of GaSe-PC cavity is changed in its wavelength, the densities of photon states at different wavelengths would be described by a Lorentzian function ($f_{Lorentzian}$) obtained from the fitting of $Mode_2$ in Fig. 1(d). The cavity-enhanced SFG is the mixing of $Mode_1$ and $Mode_2$. Therefore, when the pump laser is scanned across the $Mode_2$, the SFG power would vary in the form of $Mode_2$'s Lorentzian lineshape

$f_{Lorentzian}$, as indicated by the fitting curve in Fig. 3(b). In the $SHG_2$ pumped by $Mode_2$ itself, to generate one frequency-upconverted photon, there are two photons of the fundamental wave participate. Considering the photon density of the fundamental wave is determined by $Mode_2$'s Lorentzian function, $SHG_2$ spectrum shown in Fig. 3(b) is fitted well using the squared $Mode_2$'s Lorentzian function $(f_{Lorentzian})^2$. Governed by the same principle of photon-participation and conservation processes, the spectra of $THG_2$, $CSFG_1$ and $CSFG_2$ present lineshapes described by cubic, linear and squared $Mode_2$'s Lorentzian function, respectively.

**4. Conclusion**

In conclusion, we have demonstrated the achievements of multiple OFCs in 2D material with the integration of a silicon PC cavity. Enhanced by the high densities of electrical fields in the cavity modes, it is possible to realize various second-order nonlinear processes in 2D materials with a low-power CW pump. It assists efficient nonlinear mixings of two pump lasers and facilitates the deterministic frequency-conversion processes of SFG, cascaded SFG and THG. In a fabricated device, we estimated the SHGs of a five-layer GaSe are enhanced by factors around 1,000, and the obtained cascaded SFGs have comparable intensities with SHGs. To the best of our knowledge, this is the first realization of cascaded second-order nonlinear processes in 2D materials. The realizations of ultralow power CW pumped OFCs present great potentials to utilize 2D materials' distinct second-order nonlinearities in nonlinear optoelectronic applications, including linear electro-optic modulator, laser frequency converters, frequency-upconversion detectors.[41] In addition, GaSe-assisted strong OFCs on a silicon PC cavity indicate that 2D materials with high $\chi^{(2)}$ could function as a new type of chip-integrated active materials to develop high-performance nonlinear devices for future on-chip interconnects.

**Supporting Information**

Fabrication of the integrated GaSe-PC cavity; Growth recipes of the single-crystal GaSe; Characterizations of GaSe's crystal structure; Experimental arrangement; Raman spectroscopy of the GaSe flake integrated on the PC cavity; Estimations of the SHG enhancement factors by the PC cavity; Frequency-upconversion signals from the bare silicon PC cavity; Spatial mappings of the OFC signals

**Acknowledgements**

Financial support was provided by the NSFC (61522507, 61775183, 11634010, 11761131001), the Key Research and Development Program (2017YFA0303800), the Key Research and Development Program in Shaanxi Province of China (2017KJXX-12), and the Fundamental Research Funds for the Central Universities (3102017jc01001). We thank the AFM measurements by the Analytical & Testing Center of NPU.

**Supporting Information**

**Table of contents:**

**A. Fabrication of the integrated GaSe-PC cavity**

We fabricate PC cavities on a silicon-on-insulator (SOI) wafer with a 220 nm thick top silicon slab. The PC cavity patterns are defined on a spin-coated electron-beam resist using the electron beam lithography, which are then transferred into the top silicon layer with an inductively coupled plasma etching. After removing the residual resist layer, the bottom buried oxide layer is removed by a wet chemical etching using a diluted hydrofluoric acid to air-suspend the PC cavities. To obtain a PC cavity with resonant modes at the telecom-band of the employed tunable laser, the lattice constant and radii of the periodic air-holes are designed as $a = 450$ nm and $r = 0.25a$, respectively. The cavity defect is formed by outward shifting the central four air-holes vertically and horizontally with a distances of 0.2$a$ and 0.12$a$, respectively, which ensures an ultrasmall $V_{mode}$ of the resonant modes.[1] To facilitate its mode-coupling with the used vertical microscope, part of the air-holes around the central defect are shrunk to a radius of 0.2$a$ to shift the resonant mode into an asymmetric distribution.[2] Simulated with a finite element method (COMSOL Multiphysics software), the designed cavities could have quality ($Q$) factors as high as 9,000. Unfortunately, due to the fabrication imperfections, the measured $Q$ factors of the fabricated PC cavities are around 3,000.

The bulk GaSe single-crystal is grown by the Bridgman method, as detailed in the following part. To assist its integration with a PC cavity, few-layer GaSe flakes are mechanically exfoliated onto a polydimethylsiloxane stamp using the Scotch tape, which are then dry transferred onto the cavity with the assistance of a micromanipulation system.[3] The high quality GaSe crystal enables the easy exfoliation of large few-layer GaSe flakes with dimensions over 1000 $\mu m^2$ to cover the cavity completely and precisely.

**B. Growth recipes of the single-crystal GaSe[4]**

To grow GaSe single-crystal, we used high-purity Se (6N, 99.9999%) and Ga (6N, 99.9999%) from Emei Corp., Ltd. (Emei, China) as the raw materials. A fused quartz ampoule coated with a carbon film was used to prevent adhesions. Ga source was baked at 673 K for 4 hours under high vacuum to remove the oxidization layer. Ga and Se were mixed in a stoichiometric ratio and then sealed at $10^{-5}$ Torr. A single-temperature zone rocking furnace was used for the synthesis of the polycrystals.

A two-temperature zone furnace was then used to grow the GaSe single crystal by the vertical Bridgman method. The upper and lower zones were set at 1293 K and 1173 K, separately, and the temperature gradient was 10 K/cm. The crystal growth rate was 0.5 mm/h. A BN-crucible with an inner diameter of 22 mm was used to grow the GaSe crystal.

## C. Characterizations of GaSe's crystal structure[4]

The phase and crystal structure of the as-grown GaSe crystal were identified by powder X-ray diffraction and transmission electron microscopy (TEM), which have been partially reported in our recently published work.[4] Figure S1(a) shows the powder diffraction pattern of the GaSe crystal, which is well consistent with the theoretical curve of GaSe obtained from JCPDS: 37-0931, as shown in Fig. S1(b). It confirms that the hexagonal GaSe crystal is grown with cell parameters of $a=b=3.749$ Å, and $c=15.907$ Å, space group $D_{3h}^1$. Usually, the symmetry of the X-ray rocking curve is the direct evidence of crystal's structural uniformity. We display the X-ray rocking curve of the (004) face of the as-grown GaSe crystal in Fig. S1(c). The peak shape is symmetric with a FWHM about 46 arcs, which is the smallest value ever reported for GaSe crystals, which are 0.15° in Ref. [5], 0.07° in Ref. [6], 0.04° in Ref. [7], and split peaks in Ref. [8].

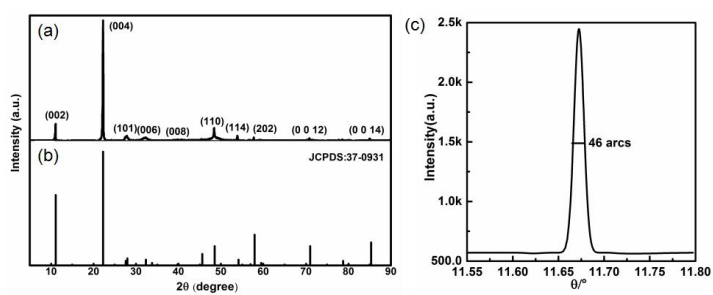

**Figure S1**. XRD measurements of the as-grown GaSe crystal[4]. (a) The powder X-ray diffraction pattern (GaSe) of the GaSe sample. (b) The database of GaSe from JCPDS: 37-0931. (c) The rocking curve of the (004) face.

To further examine the crystal structure and quality of the as-grown GaSe crystal, high-resolution TEM measurements are implemented using a FEI Talos F200X operated at 200 kV. The exfoliated GaSe flakes were transferred onto a TEM copper grid using a dry transfer method. The measurement results are shown in Fig. S2. The GaSe flake is verified in the low-magnification mode, as shown in Figs. S2(a). And the high-resolution TEM images of the dashed box area in (a) are shown in Figs. S2(b) and (c). The crystalline lattice can be clearly resolved, with the in-plane lattice constant of GaSe ~ 0.38 nm, belonging to the ε-GaSe polytype with the 2H (A+B) stacking feature, which is also predicted to be the stable polytype in bulk. The selected area electron diffraction (SAED) pattern with incident zone axis <001> is shown in Fig. S2(d). A typical six-fold symmetry presents, revealing the good crystalline quality of the hexagonal GaSe lattice. Then, according to Figs. S2(e) and (f), the energy dispersive X-ray spectrum (EDX) mappings confirm the chemical compositions of Ga and Se homogeneously distributed over the GaSe flake. The atomic ratio calculated from the EDX result is Ga:Se ≈ 1:1.02, confirming the monochalcogenide phase of the grown GaSe in the Bridgman technique.

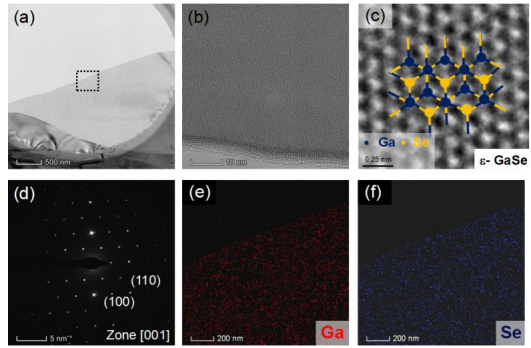

**Figure S2**. TEM characterizations of a few-layer GaSe flake. (a) TEM image with a low-magnifications. (b, c) High-resolution TEM images with the schematic of ε-GaSe lattices with the 2H-type stacking. (d) SAED pattern of a GaSe flake, which is taken with a beam parallel to the <001>. (e, f) EDX elemental mappings of Ga and Se.

## D. Experimental arrangement

The experimental measurements over the GaSe-PC cavity are implemented using a vertically coupled cross-polarization microscope,[9] as schematically shown in the Fig. 1(c) of maintext. The objective lens (OL) of the microscope is a near-infrared (NIR) anti-reflective one with a 50× magnification and a numerical aperture of 0.42. A white light source and a silicon CCD are employed to image the GaSe-PC cavity's position during the measurements, which are not shown in Fig. 1(c) of the maintext.

To examine resonant peaks of the PC cavity, the cavity reflection is measured with an external excitation light source. A polarized beam splitter (PBS) is employed in the microscope system to achieve orthogonally polarized excitation laser and collection signal, which allows a high signal to noise ratio of the cavity mode. A half wave plate (HWP) is inserted between the PBS and OL to control the direction of laser polarization with respect to the PC cavity's orientation. The employed excitation source is a narrowband tunable CW laser (Yenista, T100S-HP/CL), which is incident from the *Input*1 port. The cavity reflection is vertically collected by the same OL and then reflected by the PBS into a telecom photodiode at the *Output*1 port. By tuning the laser wavelength around the resonant peaks with a step of 0.005 nm, cavity's reflection spectra are recorded, as shown in Fig. 1(d) of the maintext.

OFC measurements from the GaSe-PC cavity are carried out in the same vertical microscope. To mix two pump lasers into the GaSe-PC cavity, another tunable laser is incident from the *Input*2 port, which is controlled by a flip mirror (FM) in the signal collection path to replace the *Output1* port. The OFC signals scattered from the PC cavity are collected by the OL, which are then reflected by a longpass (cut-off wavelength of 1,000 nm) dichroic mirror (DM) placed in the signal collection path. The OFC spectra are examined by a 0.5 m spectrometer mounted with a cooled silicon camera or a photomultiplier tube (PMT). To evaluate the cavity enhancement factor over the SHG, a pulsed picosecond laser is employed as well (PriTel FFL-20MHz).

**E. Raman spectroscopy of the GaSe flake integrated on the PC cavity**

The ultrathin GaSe flakes have low stability in air. According to our previous testing, a ~10 nm thick GaSe sheet will completely degrade in air after 3 months, while this behavior is more stable than few-layer black phosphorus flakes. To confirm that the GaSe flake integrated on the PC cavity has no great degradation after the device fabrication and OFC measurements, we carry out the Raman spectroscopic studies. A home-built Raman spectroscopy system is employed with an excitation laser at 532 nm, which has an objective lens with a numerical aperture of 0.42 to focus the laser and collect the Raman scattering signal. The pump power is controlled to be lower than 10 µW to avoid the laser-induced degradation. After filtering the pump laser using two long-pass filters, the Raman signal is monitored and analyzed by a spectrometer mounted with a cooled silicon camera.

Figure S3 shows the measured Raman spectra of the GaSe flake after its integration with the PC cavity. Three Raman scattering peaks are obtained at the 135, 214, and 308 cm$^{-1}$, which are consistent with those reported in Ref. [10]. It verifies the dry transfer process of the GaSe flake could maintain its quality. After the whole OFC measurements with a pump of the CW laser around 1550 nm, we carried out the Raman measurement again, as shown in Fig. S3. The Raman signals do not show obvious variations. However, we observe the 532 nm excitation laser changes the Raman signal remarkably, while it only illuminates the GaSe flake no more than 1 minute. This could be attributed to the effective absorption of the 532 nm laser by the GaSe flake considering its optical bandgap around 600 nm. For the illumination of CW laser around 1550 nm, the GaSe flake is stable since the absence of light-absorption. The OFC signals are monitored as well to examine the degradation process of the GaSe flake, showing no observable variations after each measurement. In addition, the device is stored in a vacuum box after each measurement to extend its lifetime.

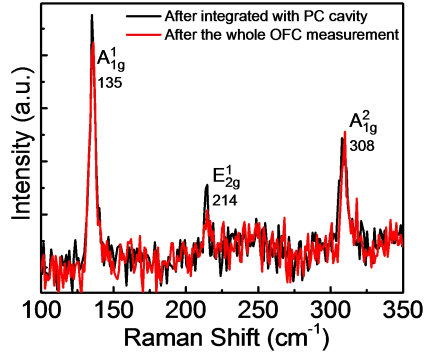

**Figure S3**. Raman scattering spectra of the GaSe flake after its integration with the PC cavity and after the whole OFC measurements.

## F. Estimations of the SHG enhancement factors by the PC cavity

Due to the optical losses from various optics of the measurement setup as well as the optical absorption by the silicon substrate, it is complicated to measure the absolute SHG power from the GaSe-PC cavity. To evaluate the SHG enhancement factors by the PC cavity, we switch the CW laser into a pulsed laser to pump the GaSe-PC cavity. The pulsed laser has a wavelength (at 1560 nm) off-resonant from the cavity modes. In this regime, both the on-resonance CW pump and the off-resonance pulsed pump share the same excitation and collection optical paths, as well as the same location of GaSe layer to maintain the dielectric environment of the SHG emission. The generated SHG signal is measured using a PMT.

The pulsed laser has Gaussian-function pulses with a width of $\tau = 8.8$ ps and a repetition rate of $Rep = 18.5$ MHz. Assuming the Gaussian pulse has a complex envelope with constant phase and Gaussian magnitude, the time-function of the pulse power could be written as:

$$P(t) = P_0 \exp(-2t^2/\tau^2) \tag{S1}$$

where $P_0$ is the peak power of a single pulse. This peak power $P_0$ could be calculated from the measured averaged power $P_{ave}$ by

$$P_0 = (P_{ave} \times 1\text{s}) / (Rep \times \int P(t)\, dt) \qquad (S2)$$

In the second-order nonlinear process, the SHG power $P_{SHG}$ is proportional to the squared pump power $P_{pump}$, i.e., $P_{SHG} \propto P^2_{pump}$, and for a pulse laser it should be proportional to $P^2(t)$. For the employed pulsed laser, in a time duration of one second, the energy of SHG is proportional to $Rep \times \int P^2(t)\, dt$.

In the experiment, the off-resonance pulsed laser is first incident on the GaSe-PC cavity, whose averaged power focused on the device is measured as $P_{ave} = 2.9$ mW. The SHG power measured by the PMT is about 35 nW. Then, the pulsed laser is switched into the on-resonance CW lasers at 1528.0 nm and 1587.6 nm. We measure the required powers of the two on-resonance CW lasers when the generated SHGs have powers of 35 nW in the PMT, which are 2.8 mW and 3.9 mW for the lasers at 1528.0 nm and 1587.6 nm. By further considering the far-field coupling-in efficiencies, the powers coupled into the $Mode_1$ and $Mode_2$ of PC cavity are 0.13 mW and 0.24 mW, respectively.

Combining with Eqs. (S1)-(S2) and $P_{ave} = 2.9$ mW, the SHG energy generated by the pulsed laser in one second equals to that pumped by a CW laser with a power of $\sqrt{Rep \times \int P^2(t)\, dt}\, / 1\text{s} = 171.4$ mW. We therefore calculate the SHG enhancement factors by $Mode_1$ and $Mode_2$ of the PC cavity are 1,320 and 714, respectively. Here, the factors of the focused pulsed laser spot area and the in plane areas of the cavity modes are negligible, which have the similar values.

## G. Frequency-upconversion signals from the bare silicon PC cavity

Before the integration of the few-layer GeSe, the cavity-enhanced frequency-upconversion signals from the bare silicon PC cavity are examined as well. Two CW lasers at the wavelengths of 1523.7 nm and 1583.3 nm are incident from the *Input1* and *Input2* ports of the experiment setup, respectively, which are then resonantly coupled into the $Mode_1$ and $Mode_2$ of the bare PC cavity. Silicon has a high third-order nonlinearity in the telecom-band. The cavity-enhanced electrical field in the PC cavity could efficiently couple with the silicon slab to enable third-order nonlinear processes.[11] Frequency-upconversion spectrum of the bare PC cavity is acquired, as displayed in Fig. S4. Different from the OFC spectrum of the GaSe-PC cavity, only two peaks at the wavelengths of 507.9 nm and 527.8 nm are obtained from the bare PC cavity, which correspond to the third harmonic generations (THGs) of the two pump lasers. Also, even when the pump powers of the two CW lasers are both increased to 100 mW, the THGs from the bare PC cavity are much weaker than those obtained from the GaSe-PC cavity. These results indicate the observed SHGs, SFGs, cascaded SFGs from the GaSe-PC cavity are attributed to the second-order nonlinearity of the GaSe flake.

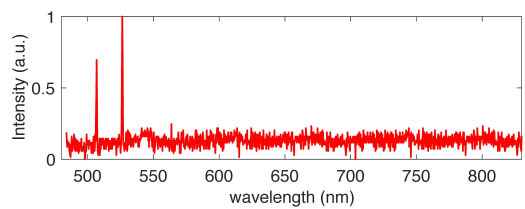

**Figure S4.** Frequency-upconversion spectrum obtained from the bare silicon PC cavity when pumped by two on-resonance lasers.

## H. Spatial mappings of the OFC signals

The cavity-enhanced OFCs are proved as well using spatial mappings of the SHG and SFG signals with on-resonance pumps, as shown in Fig. S5. The device is mounted on a two-dimensional high-resolution stage with a moving step of 100 nm. The generated OFC signals are monitored using a monochrometer mounted with a photomultiplier tube (PMT). Specific OFC signal could be filtered out from other OFC signals using the monochrometer, whose spatial distribution is then measured by the PMT. Because the cavity-enhanced OFCs are pumped by the evanescent field of cavity's resonant mode, efficient OFCs can only be observed around the cavity defect region. Figures S5(a), (b), and (c) display the spatial mapping results of $SHG_1$, $SHG_2$, and $SFG$, respectively. Here, since the cascaded SFGs are not so strong, it is difficult to obtain their clear spatial mappings. Strong $SHG_1$ and $SHG_2$ signals are observed in an area of 2.5×2 μm$^2$, which are consistent with the simulated electrical field distributions of the resonant $Mode_1$ and $Mode_2$ shown in Figs. 2(a) and (b) of the maintext. $SFG$ has a smaller area than those of the SHG, which is determined by the multiplications of the electrical field distributions of $Mode_1$ and $Mode_2$. Outside the cavity region, the GaSe layer is only pumped by the vertically illuminated laser, which is too weak to yield observed OFCs for the CW pump.

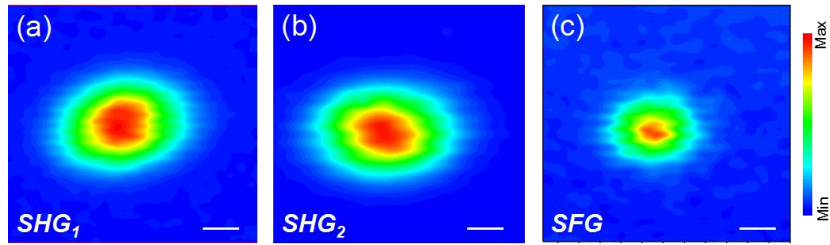

**Figure S5**. Spatial mappings of the $SHG_1$, $SHG_2$, and $SFG$. Scale bar: 1 μm.